\documentclass[aps,pra,reprint,superscriptaddress]{revtex4-2}

\usepackage[hiresbb]{graphicx}

\usepackage{times}
\usepackage{mathrsfs}

\usepackage{amsmath}
\usepackage{amssymb}

\usepackage{mathtools}
\usepackage[inline]{enumitem}

\newcommand{\defeq}{\coloneqq}

\newcommand{\lsim}[1]{\sim_{#1}}

\newcommand{\abs}[1]{\lvert #1 \rvert}

\newcommand{\lnorm}[2]{\lVert #1 \rVert_{#2}}

\newcommand{\lev}[2]{\langle #1 \rangle_{\hspace{-0.1em}#2}}

\newcommand{\dlev}[2]{\left\langle #1 \right\rangle_{\hspace{-0.4em}#2}}

\newcommand{\inpr}[2]{\langle #1, #2 \rangle}

\newcommand{\linpr}[3]{\langle #1, #2 \rangle_{\hspace{-0.1em}#3}}

\newcommand{\Star}[1]{#1\ensuremath{^*}\kern-\scriptspace}
\newcommand{\CStar}{\Star{\ensuremath{\mathrm{C}}}}

\newcommand{\hsp}[1]{\mathcal{#1}}

\newcommand{\tr}[1]{\mathrm{Tr}[#1]}
\newcommand{\dtr}[1]{\mathrm{Tr}\left[#1\right]}

\newcommand{\Id}{\mathrm{Id}}

\newcommand{\dcomm}[3]{\dlev{\frac{[#1,#2]}{2i}}{#3}}
\newcommand{\dacomm}[3]{\dlev{\frac{\{#1,#2\}}{2}}{#3}}

\newcommand{\ob}[1]{\mathcal{#1}}

\newcommand{\obsp}[1]{\mathcal{#1}}
\newcommand{\obspc}[1]{R(#1)}
\newcommand{\obspq}[1]{S(\hsp{#1})}

\newcommand{\stsp}[1]{\mathcal{#1}}
\newcommand{\stspc}[1]{W(#1)}
\newcommand{\stspq}[1]{Z(\hsp{#1})}

\newcommand{\lobsp}[2]{\obsp{#1}_{\hspace{-0.1em}#2}}
\newcommand{\lobspc}[2]{R_{#2}(#1)}
\newcommand{\lobspq}[2]{S_{\hspace{-0.1em}#2}(\hsp{#1})}

\newcommand{\Mrho}{M\hspace{-0.1em}\rho}
\newcommand{\Nrho}{N\hspace{-0.1em}\rho}
\newcommand{\Jrho}{J\hspace{-0.1em}\rho}

\newcommand{\process}[1]{\mathcal{#1}}

\newcommand{\pb}[2]{#1_{\!#2}^{\ast}}
\newcommand{\pf}[2]{#1_{\hspace{-0.1em}#2\ast}}

\newcommand{\dpf}[2]{#1_{\!#2\ast}^{\vphantom{\hspace{-0.5pt}\ast}}}

\newcommand{\spb}[1]{#1^{\hspace{-0.5pt}\ast}}
\newcommand{\spf}[1]{#1_{\hspace{-0.5pt}\ast}}

\newcommand{\err}[3]{\varepsilon_{\hspace{-0.1em}#3}(#1;#2)}
\newcommand{\gerr}[4]{\varepsilon_{\hspace{-0.1em}#4}(#1, #2;#3)}

\newcommand{\dst}[3]{\eta_{#3}(#1;#2)}
\newcommand{\gdst}[4]{\eta_{#4}(#1, #2;#3)}

\newcommand{\loss}[3]{\lambda_{#3}(#1;#2)}
\newcommand{\gloss}[4]{\lambda_{#4}(#1, #2;#3)}

\begin{document}

\title{A Universal Formulation of Uncertainty Relation for Error and Disturbance}

\author{Jaeha Lee}
\email[]{lee@iis.u-tokyo.ac.jp}
\affiliation{Institute of Industrial Science, The University of Tokyo, Chiba 277-8574, Japan.}

\author{Izumi Tsutsui}
\email[]{izumi.tsutsui@kek.jp}
\affiliation{Theory Center, Institute of Particle and Nuclear Studies, High Energy Accelerator Research Organization (KEK), Ibaraki 305-0801, Japan.}

\begin{abstract}
We present a universal formulation of uncertainty relation valid for any conceivable quantum measurement and the resultant observation (observer) effect of statistical nature.  Owing to its simplicity and operational tangibility, our general relation is also experimentally verifiable.  Our relation violates the traditional na{\"i}ve bound $\hbar/2$ for the position-momentum measurement while respecting Heisenberg's original philosophy of the uncertainty principle.  Our error-disturbance relation admits a parallel formulation to our relation for errors, which also ebmraces the standard Kennard--Robertson (Schr{\"o}dinger) relation as a special case; this attains a unified picture of the three orthodox realms of uncertainty regarding quantum indeterminacy, measurement, and observation effect within a single framework.
\end{abstract}


\maketitle


\section{Introduction\label{sec:intro}}

In recent years, we have seen a consistent --- some are rapid while others are more steady --- progress of quantum information technologies, and it should now be evident that their developments in coming years shall come to be one of the most basic foundations upon which our future society thrives.   All of these technologies are made available by the application of quantum mechanics, which was established nearly a century ago, but yet defies our deeper understanding in many respects, such as the non-local correlation arising from quantum entanglement and the non-causal change inherent in quantum measurement.  The crux of the matter behind these phenomena is arguably the renunciation of local reality, and in this regard the uncertainty principle has been deemed as the basis to guide us to the genuine comprehension of the quantum world.  Nevertheless, we have also been aware that the uncertainty principle, which was originally introduced by Heisenberg \cite{Heisenberg_1927}, is too vague to deduce rigorously testable statements, except for providing intuitive arguments which are oftentimes helpful yet could occasionally be deceptive and misleading.

The earliest attempt to remove the vagueness was made by Kennard \cite{Kennard_1927} who provided a mathematical formulation of the uncertainty principle in terms of the standard deviations for the pair of position and momentum observables, giving the familiar lower bound $\hbar/2$ for their product.  Subsequently, its generalization for arbitrary observables $A$ and $B$ was given by Robertson \cite{Robertson_1929} with the lower bound expressed by the expectation value of their commutator $\abs{ \lev{ [A, B] }{\rho} }/2$ for the state $\rho$ under consideration.  On account of the mathematical clarity, the Kennard--Robertson inequality became a standard textbook material as an exposition of the uncertainty principle, despite the fact that it has little to do with the notion of measurement to which Heisenberg attributed the cause of the uncertainty in his discourse.

In fact, as typically exemplified by his famous gamma-ray microscope Gedankenexperiment, Heisenberg did entertain the concept of measurement for devising his uncertainty principle, where he considered the error in the position measurement and the subsequent disturbance occurred in the momentum and argued that the product of the two has a lower bound of the order of the Planck constant.  It was thus clear that this line of thought, which captures the essence of the quantum \lq indeterminatenes\rq\ \cite{Heisenberg_1930}, should be followed in order to establish a more genuine uncertainty relation that governs the quantum phenomena.

Among the various attempts \cite{Yuen_1973, Werner_2004, Miyadera_2008, Watanabe_2011, Watanabe_2011_06, Busch_2013} to explicitly incorporate measurement in the formulation of uncertainty relations, one of the most recognized achievement has been Ozawa's work \cite{Ozawa_2003, Ozawa_2004_01}, which adopts the indirect measurement scheme after Arthurs, Kelly, and Goodman \cite{Arthurs_1965, Arthurs_1988}.  Here, one considers the system of an ancillary meter device in addition to the quantum system of interest, thereby realizing the concepts of error and disturbance in concrete terms.   His uncertainty relation, obtained for an arbitrary pair of self-adjoint observables $A$ and $B$, contains not only the product term of the error and the disturbance but also two other terms involving the standard deviations.  Whereas these additional terms offer the possibility of \lq breaking\rq\ the lower bound espoused by Heisenberg, it is argued \cite{Werner_2019} that they may obscure Heisenberg's original spirit instilled in his principle.  Another notable shortcoming \cite{Werner_2004, Koshino_2005} often pointed out is that his relation entails objects that are generally unobtainable from the measurement outcomes, which is considered to be a major impediment to its experimental verification except for narrow cases in which certain specific properties of the system could be exploited \cite{Erhart_2012}.

In this paper, we present a novel uncertainty relation for error and disturbance associated with measurement in a conceivably most universal framework, following our earlier works \cite{Lee_2020_01, Lee_2020_01_Entropy} on the uncertainty relation for errors, where (part of) the formulation is adopted.  In particular, our framework does not require any model of measurements such as the indirect measurement scheme, which entails completely positive quantum-state transformations, and deals directly with the measurement outcomes and transformed states, thereby providing us with a clear understanding of the various manifestations of quantum uncertainty in a unified manner. The universality also ensures the general validity of the resultant uncertainty relations for any measurements including the familiar positive-operator valued measure (POVM) measurements now extensively used in quantum information, where one is to adopt Kolmogorov's measure-theoretic formalism \cite{Kolmogorov_1933} to model probability.  Besides, our formulation allows us to focus exclusively on the product of error and disturbance to make the uncertainty relation perhaps more in line with Heisenberg's original idea.

This paper is organized as follows.  After this introduction, in Sec.~\ref{sec:setup}, we summarize the basic objects of our geometric framework.  The subsequent three sections are devoted to the explanation of the necessary tools for our formulation, in which general processes and its implications are investigated.  Our error and disturbance, or loss in general, are defined in Sec.~\ref{sec:loss_process}, followed by an exposition of its properties regarding process compositions in Sec.~\ref{sec:composition_law}.  The key observation as to how the tradeoff relation between the errors of two measurements can arise is revealed in Sec.~\ref{sec:ur_joint}.  Our main result, the uncertainty relation for error and disturbance, is then presented in Sec.~\ref{sec:ur}, and its affinity with Heisenberg's original idea is discussed in Sec.~\ref{sec:affinity}.  Finally, we provide our conclusion and discussions in Sec.~\ref{sec:discussion}.

\section{Systems\label{sec:setup}}

Let us first briefly recall the basic objects of the geometric framework proposed in our previous papers \cite{Lee_2020_01, Lee_2020_01_Entropy} that form the basis of our framework.

Let $\stspq{H}$ and $\obspq{H}$ respectively denote the state space and the observable space of a quantum system $\hsp{H}$.  For the purpose of this paper, let them be respectively modelled with the convex set of all the density operators and the linear space of all the self-adjoint operators on the Hilbert space $\hsp{H}$ associated with the system.  For an observable $A \in \obspq{H}$, each quantum state $\rho \in \stspq{H}$ furnishes a seminorm $\lnorm{A}{\rho} \defeq \sqrt{ \lev{A^{\dagger}A}{\rho} }$ on $\obspq{H}$ inherited from the one defined for general linear operators, where an operator with the superscript dagger denotes its Hilbert-space adjoint.  Here, we have introduced 
\begin{equation}\label{def:ltrace}
\lev{X}{\rho}
	\defeq \tr{X\rho}
\end{equation}
for any pair of a linear operator $X$ on $\hsp{H}$ and $\rho \in \stspq{H}$.  The seminorm induces an equivalence relation ${A \lsim{\rho} B} \iff {\lnorm{ A - B }{\rho} = 0}$ over $\obspq{H}$, which results in the partitioning of the observables into equivalence classes.  These equivalence classes of quantum observables collectively form a quotient space $\obspq{H}/{\lsim{\rho}}$, the completion of which we denote by $\lobspq{H}{\rho}$ and term it the \emph{space of local quantum observables} at $\rho$; this can be visualized as a `tangent space' attached to the point $\rho \in \stspq{H}$ of the state space.  As commonly practiced, we make an abuse of notation to denote the equivalence class with one of its representatives.  The space $\lobspq{H}{\rho}$ bears a unique inner product $\linpr{A}{B}{\rho} \defeq \lev{ \{A,B\} }{\rho}/2$ characterized by the anti-commutator $\{A,B\} \defeq AB + BA$ of observables, which is compatible with the quotient norm in the sense of $\lnorm{A}{\rho}^{2} = \linpr{A}{A}{\rho}$.  We call the resultant bundle consisting of the base space $\stspq{H}$ with the fibres $\lobspq{H}{\rho}$ attached to each of the points $\rho \in \stspq{H}$ the \emph{bundle of quantum-observable spaces} (or simply, the \emph{quantum bundle}) in this paper.

As for the classical counterpart, let $\stspc{\Omega}$ and $\obspc{\Omega}$ respectively denote the state space and the observable space of a classical system, which, for the purpose of this paper, shall be respectively modelled with the convex set of all the probability distributions and the linear space of all the real functions on a sample space $\Omega$.  In a parallel manner as above, a probability distribution $p \in \stspc{\Omega}$ furnishes a seminorm $\lnorm{ f }{p} \defeq \sqrt{\lev{ f^{\dagger} f }{p}}$ for each $f \in \obspc{\Omega}$ inherited from that defined for complex functions, where a function with the superscript dagger denotes its complex conjugate.  Here, we have introduced
\begin{equation}\label{def:lint}
\lev{ z }{p}
	\defeq \int_{\Omega} z(\omega)p(\omega)\,d\omega
\end{equation}
for any pair of a complex function $z$ on $\Omega$ and $p \in \stspc{\Omega}$.  As before, the seminorm induces an equivalence relation ${f \lsim{p} g} \iff {\lnorm{ f - g }{p} = 0}$ over the space $\obspc{\Omega}$ which, in turn, induces the quotient space $\obspc{\Omega}/{\lsim{p}}$.  Its completion yields a `tangent space' at each $p \in \stspc{\Omega}$ which will be denoted by $\lobspc{\Omega}{p}$ and addressed as \emph{the space of local classical observables} at $p$.  It is easy to check that the quotient norm admits a unique inner product $\linpr{ f }{ g }{p} \defeq \lev{ f g }{p}$ that satisfies $\lnorm{ f }{p}^{2} = \linpr{ f }{ f }{p}$.  We call the resultant bundle consisting of the base space $\stspc{\Omega}$ with the fibres $\lobspc{\Omega}{p}$ attached to each of the points $p \in \stspc{\Omega}$ the \emph{bundle of classical-observable spaces} (or simply, the \emph{classical bundle}) in this paper, which in the present paper will be used to represent the system of classical measurement outcomes.  

For general discussions, let $\stsp{S}$ and $\obsp{O}$ respectively denote the state space and the observable space of a generic system;  these encompass not only quantum systems ($\stsp{S} = \stspq{H}$, $\obsp{O} = \obspq{H}$) or classical systems ($\stsp{S} = \stspc{\Omega}$, $\obsp{O} = \obspc{\Omega}$) as specific examples, but also their composite systems, or even beyond those, granted that parallel constructions are possible (a rigorous study on our framework in the context of generalized probabilistic theory (GPT) consists one of our future works).  
Let $\lobsp{O}{s}$ be the space of local observables at $s \in \stsp{S}$ defined in a parallel manner as we have done above for the quantum ($\lobsp{O}{s} = \lobspq{H}{\rho}$) and classical ($\lobsp{O}{s} = \lobspc{\Omega}{p}$) cases, and define the bundle consisting of the base space $\stsp{S}$ with the fibres $\lobsp{O}{s}$ attached to each of the points $s \in \stsp{S}$, which admit parallel names (\textit{e.g.}, C-Q (composite) bundles, and so on) in the same vein as above.

\section{Processes\label{sec:processes}}

Let us next introduce the general notion of \emph{processes} (or, transformations), which are defined as affine maps $\process{T}: \stsp{S}_{1} \to \stsp{S}_{2}$ between state spaces $\stsp{S}_{1}$, $\stsp{S}_{2}$ of (possibly different) systems in generic terms.

Processes from quantum-state to classical-state spaces $M : \stspq{H} \to \stspc{\Omega}$ and processes between quantum-state spaces $\Theta : \stspq{H} \to \stspq{K}$ are two of the most relevant examples of processes to this paper, for the purpose of which, the former quantum-to-classical (Q-C) process, \textit{i.e.}, the case where $\stsp{S}_{1} = \stspq{H}$ and $\stsp{S}_{2} = \stspc{\Omega}$, shall be referred to as a \textit{quantum measurement}, whereas the latter quantum-to-quantum (Q-Q) process, \textit{i.e.}, the case where $\stsp{S}_{1} = \stspq{H}$ and $\stsp{S}_{2} = \stspq{K}$, shall be simply called a \textit{quantum process}.  Also relevant are processes $K : \stspc{\Omega_{1}} \to \stspc{\Omega_{2}}$ between classical-state spaces.  This classical-to-classical (C-C) process, which admits interpretations as, \textit{e.g.}, statistical processing of data, classical measurements performed on a classical system, as well as transformations of classical states, shall be referred to as a \textit{classical process}.  While the exposition of this paper can be conducted without its explicit use, the processes from classical-state to quantum-state spaces $Q : \stspc{\Omega} \to \stspq{H}$ are also of importance.  This  classical-to-quantum (C-Q) process admits interpretations as, \textit{e.g.}, quantum encoding of classical data, among others.  It goes without saying that the notion of processes also encompass maps between state spaces of composite systems, \textit{e.g.}, those of the composite systems of quantum and classical systems, and so on.

\begin{figure}
\includegraphics[hiresbb,clip,width=0.9\linewidth,keepaspectratio,pagebox=artbox]{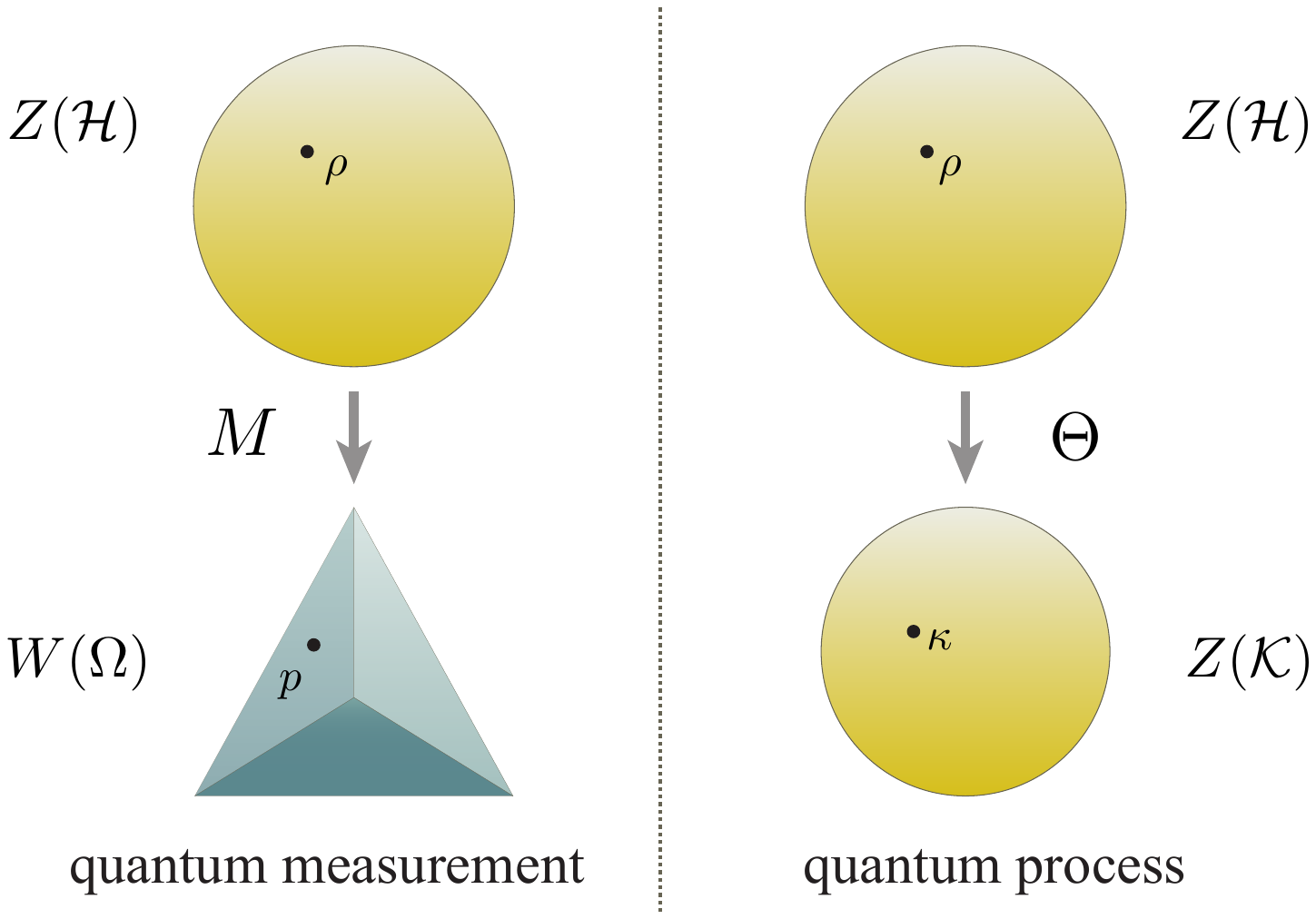}%
\caption{Our basic premise of quantum measurements and quantum processes.  The space of quantum states $\stspq{H}$ is depicted as a sphere, while the space of probability distributions $\stspc{\Omega}$ is represented by a tetrahedron.  In general, a quantum measurement can be regarded as a map $M : \stspq{H} \to \stspc{\Omega}$ associating a quantum state $\rho \in \stspq{H}$ with the output probability distribution $p \in \stspc{\Omega}$.  Similarly, a quantum process can be regarded as a map $\Theta : \stspq{H} \to \stspq{K}$ associating a quantum state $\rho \in \stspq{H}$ with the reslutant quantum state $\kappa \in \stspq{K}$ after it underwent the said process.
\label{fig:one}}
\end{figure}

The objects of primary interest of this paper are quantum measurements and its observation (observer) effects, which entail classical measurement outcomes as well as quantum-state transformations, and may thus be understood as specific examples of processes introduced here (see FIG.~\ref{fig:one}).

\section{Quantum Measurements and Processes\label{sec:measurement-and-processes}}

The archetype of quantum measurements is the familiar projection measurement associated with a quantum observable $\hat{M}$.   Assuming for simplicity that $\hat{M}$ is non-degenerate in a finite $N$-dimensional space $\hsp{H}$, the spectral decomposition $\hat{M} = \sum_{i=1}^{N} m_{i}\, |m_{i}\rangle\langle m_{i}|$ induces an affine map
\begin{equation}\label{def:projection_measurement}
M : \rho \mapsto (\Mrho)(m_{i})
	\defeq \dtr{|m_{i}\rangle\langle m_{i}| \rho},
\end{equation}
to which, in accordance with the Born rule, one may associate the interpretation that $p(m_{i}) = (\Mrho)(m_{i})$ gives the probability distribution over the sample space $\Omega \defeq \{ m_{1}, \dots, m_{N} \}$ consisting of all the eigenvalues of $\hat{M}$.  Throughout this paper, the reader may safely assume the quantum measurement $M$ to be of the form \eqref{def:projection_measurement} without missing much of the essence of the subject, although it is by no means restricted to that particular class;  as is familiar, if one is to adopt the Kolmogorovian measure-theoretic formalism of probability, a quantum measurement admits a representation by a positive-operator valued measure (POVM).

The archetype of the quantum processes of our interest is the `wave-function collapse', which is traditionally associated with the projection postulate \cite{Neumann_1932, Lueders_1951} given a projection measurement.  For the projection measurement of $\hat{M}$, the postulate dictates that the initial quantum state collapse to one of its eigenvectors $\vert m_{i}\rangle$ with probability $\abs{ \langle m_{i} | \psi \rangle }^{2}$ upon measurement;  this uniquely extends to mixed quantum states $\rho$, thereby resulting in an affine map
\begin{equation}\label{def:collapse}
\Theta : \rho \mapsto \Theta\rho
	\defeq \sum_{i=1}^{N} \dtr{|m_{i}\rangle\langle m_{i}| \rho} \cdot |m_{i}\rangle\langle m_{i}|
\end{equation}
from the quantum-state space $\stspq{H}$ to itself.  Again, the reader may assume the quantum process $\Theta$ to be of the form \eqref{def:collapse} induced by the wave-function collapse, although our map $\Theta$ may describe state changes much more general than that.

In fact, as is stated in Sec.~\ref{sec:processes}, the sole constraint we impose on the processes $M : \stspq{H} \to \stspc{\Omega}$ and $\Theta : \stspq{H} \to \stspq{K}$ is affineness, \textit{i.e.}, the maps preserve the structure of the probabilistic mixture 
\begin{align}\label{def:affine}
M\hspace{0pt}(\lambda \rho_{1} + (1-\lambda)\rho_{2})
	&= \lambda M\hspace{0pt}\rho_{1} + (1-\lambda) M\hspace{0pt}\rho_{2}, \\
\Theta\hspace{0pt}(\lambda \rho_{1} + (1-\lambda)\rho_{2})
	&= \lambda \Theta\hspace{0pt}\rho_{1} + (1-\lambda) \Theta\hspace{0pt}\rho_{2},
\end{align} 
for $\rho_{1}, \rho_{2} \in \stspq{H}$, $0 \leq \lambda \leq 1$, which is indispensable for the self-consistent statistical interpretation of density operators, ensuring that the resultant (respectively, classical and quantum) states $\Mrho$ and $\Theta\rho$ are invariant under every (pure-state) decomposition of a quantum state $\rho$.  In other words, our $M$ and $\Theta$ effectively belongs to the broadest class of maps between the state spaces concerned describing the most general transformation of states preserving statistical nature.  As for the latter, note that it includes a wide variety of state transformations, \textit{e.g.}, unitary evolution of closed quantum systems, non-unitary evolution of open quantum systems, quantum decoherence, observation (observer) effects, quantum channels and gates;  it should be also noted that our framework does not even assume the complete positivity of the (adjoint of the) quantum process.

\section{Pushforward and Pullback of Process\label{sec:pf_pb}}

One of the core concepts of our framework is that, every process between state spaces gives rise to an adjoint pair of \emph{local} (\textit{i.e.}, state-dependent) maps between the spaces of local observables.  The local maps, termed its \emph{pullback} and \emph{pushforward}, are found to be non-expansive linear maps satisfying certain composition laws, as described below.

Without loss of generality, the above statements shall be explicated with the example of quantum processes (Q-Q processes) for concreteness (and for later use in this paper);  the readers may also be referred to our previous works \cite{Lee_2020_01,Lee_2020_01_Entropy}, in which the statements are spelled out for quantum measurements (Q-C processes) in a parallel manner.

For a quantum process $\Theta : \stspq{H} \to \stspq{K}$, an important observation is that it uniquely induces a map $\Theta^{\prime}$ between operator spaces.  This dual notion of a quantum process, termed its adjoint, is uniquely characterized by the relation
\begin{equation}\label{char:adjoint}
\lev{ \Theta^{\prime} X }{\rho}
	= \lev{ X }{(\Theta\rho)}
\end{equation}
valid for all linear operators $X$ on $\hsp{K}$ and quantum states $\rho$ on $\hsp{H}$.  The wave-function collapse \eqref{def:collapse} under the projection postulate provides a prime example, the adjoint of which can be confirmed to read
\begin{equation}\label{char:adjoint_proj}
\Theta^{\prime} : X \mapsto \Theta^{\prime} X = \sum_{i=1}^{N} \dtr{|m_{i}\rangle\langle m_{i}| X} \cdot |m_{i}\rangle\langle m_{i}|,
\end{equation}
which fulfills \eqref{char:adjoint}.  The projection postulate is convenient in that they admit concrete expressions both for the quantum process \eqref{def:collapse} and its adjoint \eqref{char:adjoint_proj} using familiar notions, which could be useful for convincing oneself of the various claims throughout this paper by means of direct computation.

Now, regarding the quantum processes $\Theta$, we have the inequality
\begin{equation}\label{ineq:adjoint}
\lnorm{ A }{(\Theta\rho)}
	\geq \lnorm{ \Theta^{\prime} A }{\rho}
\end{equation}
for any $A \in \obspq{K}$ and $\rho \in \stspq{H}$.  This can be understood as a corollary to the Kadison--Schwarz inequality \cite{Kadison_1952}, which is in a sense a generalization of the Cauchy--Schwarz inequality to \CStar-algebras.  Indeed, its application to the adjoint $\Theta^{\prime}$ yields the evaluation $\Theta^{\prime}(N^{\dagger}N) \geq (\Theta^{\prime}N)^{\dagger}(\Theta^{\prime}N)$ valid for any normal operator $N$ on $\hsp{K}$, whereby \eqref{ineq:adjoint} follows directly.

\begin{figure}
\includegraphics[hiresbb,clip,width=0.9\linewidth,keepaspectratio,pagebox=artbox]{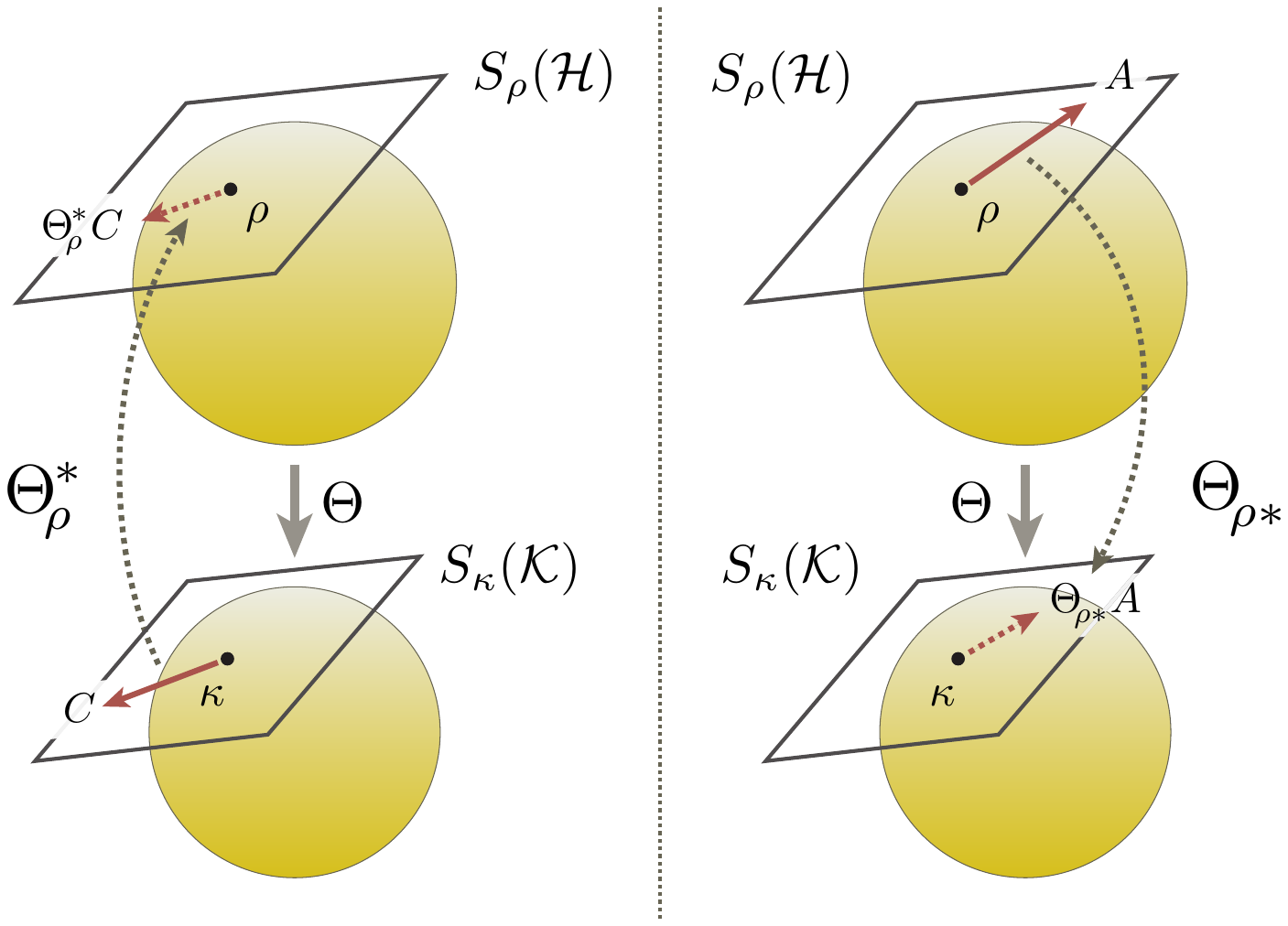}%
\caption{The pullback and the pushforward of the quantum process.  (Left) A quantum process $\Theta$ entails the pullback $\pb{\Theta}{\rho}$ from $\lobspq{K}{\kappa}$ to $\lobspq{H}{\rho}$, each of which is attached to the respective points $\kappa = \Theta\rho \in \stspq{K}$ and $\rho \in \stspq{H}$ of the corresponding state spaces.  (Right) Conversely, $\Theta$ also entails the pushforward that maps in the opposite direction.  The pullback and the pushforward are dual to each other through the relation \eqref{char:pf_and_pb_quantum-measurement}, and both of them are non-expansive maps, that is, the norm decreases (or remains unchanged) under each of the maps.
\label{figure2}}
\end{figure}

An immediate consequence of \eqref{ineq:adjoint} is the implication ${A \lsim{(\Theta\rho)} B} \implies {\Theta^{\prime}A \lsim{\rho} \Theta^{\prime}B}$.  This allows the adjoint $\Theta^{\prime}$, which was initially introduced as a map between Hilbert-space operators, to be passed to the map between their equivalence classes.  We call the resultant map
\begin{equation}\label{def:pullback}
\pb{\Theta}{\rho} : \lobspq{K}{(\Theta\rho)} \to \lobspq{H}{\rho}
\end{equation}
the \textit{pullback} of the quantum process $\Theta$ over the quantum state $\rho$ (see FIG. \ref{figure2}).  In concrete terms, this implies that, given (the equivalence class of) an operator $A \in \lobspq{K}{(\Theta\rho)}$, we have (that of) a corresponding operator $\pb{\Theta}{\rho} A \in \lobspq{H}{\rho}$.

Among the various properties of the pullback, the non-expansiveness $\lnorm{ A }{(\Theta\rho)} \geq \lnorm{ \pb{\Theta}{\rho} A }{\rho}$, which follows directly by construction, and the composition law described below are of our particular interest.  Let $\Theta : \stspq{H} \to \stspq{K}$ and $\Phi : \stspq{K} \to \stspq{L}$ be two quantum processes, with $\hsp{H}$, $\hsp{K}$, and $\hsp{L}$ being Hilbert spaces.  Since affineness is closed under map composition, the composite map $\Phi \circ \Theta: \stspq{H} \to \stspq{L}$ is itself a quantum process.  Then, the pullback of the composite process satisfies the composition law
\begin{equation}\label{eq:composition_pullback}
\pb{(\Phi \circ \Theta)}{\rho} = \pb{\Theta}{\rho} \circ \pb{\Phi}{(\Theta\rho)},
\end{equation}
which can be readily demonstrated by means of straightforward computation.

It now remains to introduce the dual notion of the pullback, which we call the \textit{pushforward} of the quantum process $\Theta$.  The pushforward
\begin{equation}\label{def:pushforward}
\pf{\Theta}{\rho} : \lobspq{H}{\rho} \to \lobspq{K}{(\Theta\rho)}
\end{equation}
of $\Theta$ over $\rho$ is defined as the Hilbert-space adjoint of the pullback \eqref{def:pullback} regarding the inner products on the space of local observables.  More explicitly, the pushforward \eqref{def:pushforward} is uniquely characterized by the relation
\begin{equation}\label{char:pf_and_pb_quantum-measurement}
\linpr{ A }{ \pb{\Theta}{\rho}C }{\rho}
	= \linpr{ \pf{\Theta}{\rho} A }{ C }{(\Theta\rho)}
\end{equation}
valid for any choices of $A \in \lobspq{H}{\rho}$ and $C \in \lobspq{K}{(\Theta\rho)}$ on the respective systems.  Since the pullback is non-expansive, its adjoint, \textit{i.e.}, the pushforward, is also non-expansive $\lnorm{ A }{\rho} \geq \lnorm{ \pf{\Theta}{\rho} A }{\Theta\rho}$.  Equivalent to the composition law \eqref{eq:composition_pullback} of the pullback is that
\begin{equation}\label{eq:composition_pushforward}
\pf{(\Phi \circ \Theta)}{\rho}
	= \pf{\Phi}{(\Theta\rho)} \circ \pf{\Theta}{\rho}
\end{equation}
of the pushforward.

It is to be reminded that the statements described above are universal among general processes given suitable constructions, which may demonstrated through obvious parallel arguments and definitions:  in general, a process $\process{T}:\stsp{S}_{1} \to \stsp{S}_{2}$ is found (by means of its adjoint $\process{T}^{\prime}$ that takes observables $\ob{B}$ in $\obsp{O}_{2}$ to those $\process{T}^{\prime}\ob{B}$ in $\obsp{O}_{1}$) to induce a pair of dual maps $\pb{\process{T}}{s} : \lobsp{O}{\process{T}(s)} \to \lobsp{O}{s}$ and $\dpf{\process{T}}{s} : \lobsp{O}{s} \to \lobsp{O}{\process{T}(s)}$, respectively termed the pullback and the pushforward of $\process{T}$ over $s \in \stsp{S}_{1}$.  The pullback and pushforward are non-expansive linear maps, \textit{i.e.} $\lnorm{ \ob{B} }{\process{T}(s)} \geq \lnorm{ \pb{\process{T}}{s}\ob{B} }{s}$ for $\ob{B} \in \lobsp{O}{\process{T}(s)}$ and $\lnorm{ \ob{A} }{s} \geq \lnorm{ \dpf{\process{T}}{s}\ob{A} }{\process{T}(s)}$ for $\ob{A} \in \lobsp{O}{s}$, and the composition laws \eqref{eq:composition_pullback} and \eqref{eq:composition_pushforward} are valid as well.  By induction, note that the latter admits obvious generalizations to the composition $\process{T}_{n} \circ \cdots \circ \process{T}_{1} : \stsp{S}_{1} \to \stsp{S}_{n+1}$ of any number of general processes $\process{T}_{k} : \stsp{S}_{k} \to \stsp{S}_{k+1}$, $n \geq k \geq 1$:  more explicitly, the composition laws for the pullback and the pushforward respectively read
\begin{equation}\label{eq:composition_pullback_process}
\pb{(\process{T}_{n} \circ \cdots \circ \process{T}_{1})}{s}
	= \pb{(\process{T}_{1})}{s_{1}} \circ \cdots \circ \pb{(\process{T}_{n})}{s_{n} }
\end{equation}
and
\begin{equation}\label{eq:composition_pushforward_process}
\dpf{(\process{T}_{n} \circ \cdots \circ \process{T}_{1})}{s}
	= \dpf{(\process{T}_{n})}{s_{n} } \circ \cdots \circ \dpf{(\process{T}_{1})}{s_{1}}
\end{equation}
for $n \geq 1$ (for the case $n=1$, we adopt the obvious convention $\varphi_{n} \circ \cdots \circ \varphi_{1} \defeq \varphi_{1}$ regarding the composition of generic maps $\varphi_{k}$), where we have defined the symbols
\begin{equation}\label{abbr:composition_states}
s_{k+1} \defeq \process{T}_{k} s_{k}, \quad s_{1} \defeq s,
\end{equation}
inductively for $n \geq k \geq 1$.

\section{Loss associated with a Process\label{sec:loss_process}}

Armed with our geometric framework, we have introduced our definition of (quantum) error \cite{Lee_2020_01, Lee_2020_01_Entropy} by the amount of contraction
\begin{equation}\label{def:error_quantum-measurement}
\err{A}{M}{\rho}
	\defeq \sqrt{ \lnorm{ A }{\rho}^{2} - \lnorm{ \pf{M}{\rho}A }{\Mrho}^{2} }
\end{equation}
induced by the pushforward of the measurement $M$ for $A \in \lobspq{H}{\rho}$ and $\rho \in \stspq{H}$.  In a parallel manner, we introduce our definition of disturbance
\begin{equation}\label{def:disturbance_quantum-process}
\dst{A}{\Theta}{\rho}
	\defeq \sqrt{ \lnorm{ A }{\rho}^{2} - \lnorm{ \pf{\Theta}{\rho}A }{\Theta\rho}^{2}}
\end{equation}
associated with a quantum process $\Theta$ for $A \in \lobspq{H}{\rho}$ over $\rho \in \stspq{H}$.  

In view of their parallel manner of construction, the error \eqref{def:error_quantum-measurement} and disturbance \eqref{def:disturbance_quantum-process} obviously share common properties.  Indeed, just as with the case \cite{Lee_2020_01, Lee_2020_01_Entropy} of the error, non-negativity $\dst{A}{\Theta}{\rho} \geq 0$ of the disturbance is guaranteed by the non-expansiveness of the pushforward, and the absolute homogeneity $\dst{tA}{\Theta}{\rho} = \abs{ t }\, \dst{A}{\Theta}{\rho}$, $\forall t \in \mathbb{R}$, as well as the subadditivity $\dst{A}{\Theta}{\rho} + \dst{B}{\Theta}{\rho} \geq \dst{A + B}{\Theta}{\rho}$ follow by a completely parallel argument;  in other words, the error and disturbance both furnish seminorms on the space $\lobspq{H}{\rho}$ of local quantum observables.

From our general point of view, the error and disturbance are specific examples of the `loss' of generic processes $\process{T}: \stsp{S}_{1} \to \stsp{S}_{2}$, which we define as
\begin{equation}\label{def:loss_process}
\loss{\ob{A}}{\process{T}}{s}
	\defeq \sqrt{ \lnorm{ \ob{A} }{s}^{2} - \lnorm{ \dpf{\process{T}}{s}\ob{A} }{\process{T}(s)}^{2} }
\end{equation}
regarding $\ob{A} \in \lobsp{O}{s}$ and $s \in \stsp{S}_{1}$;  with this general definition, the error \eqref{def:error_quantum-measurement} is the loss of a Q-C process, whereas the disturbance \eqref{def:disturbance_quantum-process} is the loss of a Q-Q process.  The properties above described for error and disturbance are universal among the losses \eqref{def:loss_process} of generic processes $\process{T}: \stsp{S}_{1} \to \stsp{S}_{2}$, namely, they furnish seminorms on the space $\lobsp{O}{s}$ of local observables at $s \in \stsp{S}_{1}$.

The loss of a process also universally admits an operational interpretation as the minimal loss concerning the local reconstruction of an observable through the process  (\textit{cf}. discussions in Refs.~\cite{Lee_2020_01,Lee_2020_01_Entropy}).  We define the quantity
\begin{equation}\label{def:loss_gauge}
\gloss{\ob{A}}{\ob{B}}{\process{T}}{s}
	\defeq \sqrt{ \lnorm{ \ob{A} - \pb{\process{T}}{s}\ob{B} }{s}^{2} + \bigl( \lnorm{ \ob{B} }{\process{T}(s)}^{2} - \lnorm{ \pb{\process{T}}{s}\ob{B} }{s}^{2} \bigr) }
\end{equation}
for the evaluation of the precision of the reconstruction of $\ob{A} \in \lobsp{O}{s}$ by means of the pullback $\tilde{\ob{A}} = \pb{\process{T}}{s}\ob{B}$ of the local observable $\ob{B} \in \lobsp{O}{\process{T}(s)}$.  This admits an interpretation as another gauge of loss, and shall be called the \emph{loss with respect to} $\ob{B}$ (\textit{abbr.} $\ob{B}$-\emph{loss}) in this paper;  the first term inside the square-root evaluates the (algebraic) deviation of the reconstruction, whereas the second term refers to the suboptimality of the choice of $\process{T}$ and $\ob{B}$ that reproduces the same $\tilde{\ob{A}}$ through the pullback.  
Specific examples include, among others, that for quantum measurements (\textit{i.e.}, Q-C processes) $\process{T} = M : \stspq{H} \to \stspc{\Omega}$, which shall be called the \emph{error with respect to} $f$ (\textit{abbr}. $f$-\emph{error}) in this paper as in Refs.~\cite{Lee_2020_01,Lee_2020_01_Entropy} and reads $\gerr{A}{f}{M}{\rho} \defeq \sqrt{ \lnorm{ A - \pb{M}{\rho}f }{\rho}^{2} + ( \lnorm{ f }{\Mrho}^{2} - \lnorm{ \pb{M}{\rho}f }{\rho}^{2}) }$ for the state $s = \rho \in \stspq{H}$ and the pair of local observables $\ob{A} = A \in \lobspq{H}{\rho}$ and $\ob{B} = f \in \lobspc{\Omega}{\Mrho}$, as well as that for quantum processes (\textit{i.e.}, Q-Q processes) $\process{T} = \Theta : \stspq{H} \to \stspq{K}$, which shall be called the \emph{disturbance with respect to} $B$ (\textit{abbr}. $B$-\emph{disturbance}) in this paper and reads $\gdst{A}{B}{\Theta}{\rho} \defeq \sqrt{ \lnorm{ A - \pb{\Theta}{\rho}B }{\rho}^{2} + ( \lnorm{ B }{\Theta\rho}^{2} - \lnorm{ \pb{\Theta}{\rho}B }{\rho}^{2}) }$ for the state $s = \rho \in \stspq{H}$ and the pair of local observables $\ob{A} = A \in \lobspq{H}{\rho}$ and $\ob{B} = B \in \lobspq{K}{(\Theta\rho)}$.  The $\ob{B}$-losses \eqref{def:loss_gauge} of other types of processes, including those of C-C processes and C-Q processes, are also spelled out in an obvious parallel manner.  One then finds that the square of the $\ob{B}$-loss admits the decomposition
\begin{equation}
\gloss{\ob{A}}{\ob{B}}{\process{T}}{s}^{2}
	= \loss{\ob{A}}{\process{T}}{s}^{2} + \lnorm{\dpf{T}{s}\ob{A} - \ob{B} }{\process{T}(s)}^{2}
\end{equation}
into the sum of the squares of the loss \eqref{def:loss_process} and the term corresponding to the suboptimality of the choice of $\ob{B}$, thereby pointing to an operational characterization of the loss as the minimum
\begin{equation}
\loss{\ob{A}}{\process{T}}{s}
	= \min_{\ob{B}} \gloss{\ob{A}}{\ob{B}}{\process{T}}{s}
\end{equation}
of the $\ob{B}$-losses over all the local observables $\ob{B} \in \lobsp{O}{\process{T}(s)}$, as well as the interpretation of the pushforward $\dpf{T}{s}\ob{A}$ as the unique (up to equivalence) optimal one that realizes it.

\section{The (De-\nobreak\hspace{0pt})\nobreak\hspace{0pt}composition Law of Loss\label{sec:composition_law}}

Given the universality of the structure of the pullback and the pushforward, as well as the parallel manner of defining the loss among generic processes, the composition law \eqref{eq:composition_pushforward_process} of the pushforward entails the (de-\nobreak\hspace{0pt})\nobreak\hspace{0pt}composition law of the loss of a composite process $\process{T}_{n} \circ \cdots \circ \process{T}_{1}$, which reads
\begin{equation}\label{eq:composition_law_process}
\loss{\ob{A}}{\process{T}_{n} \circ \cdots \circ \process{T}_{1}}{s}^{2}
	= \sum_{k=1}^{n} \loss{\dpf{(T_{k-1})}{s} \,\ob{A}}{\process{T}_{k}}{s_{k}}^{2}
\end{equation}
for any $A \in \lobsp{O}{s}$, $s \in \stsp{S}_{1}$, and $n \geq 1$.  Here, along with the symbol \eqref{abbr:composition_states} for states, we have also introduced
\begin{equation}
T_{k+1}
	\defeq \process{T}_{k+1} \circ T_{k}, \quad T_{0} \defeq \Id,
\end{equation}
defined inductively for $n - 1 \geq k \geq 0$, where $T_{0} \defeq \Id : \stsp{S}_{1} \to \stsp{S}_{1}$ is the identity map on $\stsp{S}_{1}$;  note here that, by construction, the pullback and pushforward of the identity map $\Id : \stsp{S} \to \stsp{S}$ are both the identity map on $\lobsp{O}{s}$ for all $s \in \stsp{S}$.  This (de-\nobreak\hspace{0pt})\nobreak\hspace{0pt}composition law \eqref{eq:composition_law_process}, or the `Pythagorean theorem', for the loss of composite processes admits a simple proof by induction;  noting its trivial validity for $n=1$, if it were valid for $n \geq 1$, the expansion
\begin{align}
&\hspace{-2pt} \loss{\ob{A}}{\process{T}_{n+1} \circ \cdots \circ \process{T}_{1}}{s}^{2} \\
	&\defeq \lnorm{ \ob{A} }{s}^{2} - \lnorm{ \dpf{(T_{n+1})}{s} \ob{A} }{s_{n+2}}^{2} \notag \\
	&
	\begin{aligned}
	\phantom{:}= \lnorm{ \ob{A} }{s}^{2} &- \lnorm{ \dpf{(T_{n})}{s} \ob{A} }{s_{n+1}}^{2} \\
		&+ \lnorm{ \dpf{(T_{n})}{s} \ob{A} }{s_{n+1}}^{2} - \lnorm{ \dpf{(T_{n+1})}{s} \ob{A} }{s_{n+2}}^{2}
	\end{aligned} \\
	&\phantom{:}= \loss{\ob{A}}{\process{T}_{n} \circ \cdots \circ \process{T}_{1}}{s}^{2} + \loss{\dpf{(T_{n})}{s} \ob{A}}{\process{T}_{n+1}}{s_{n+1}}^{2}
\end{align}
points to its validity for $n+1$, where we used \eqref{eq:composition_pushforward_process} in the last equality.

As a concrete example, and for later use, let the (de-\nobreak\hspace{0pt})\nobreak\hspace{0pt}composition law \eqref{eq:composition_law_process} be spelled out for the specific case regarding the composition of a quantum measurement $\process{T}_{2} = L : \stspq{K} \to \stspc{\Omega}$ after a quantum process $\process{T}_{1} = \Theta : \stspq{H} \to \stspq{K}$: the composite process furnishes a quantum measurement on $\stspq{H}$, the loss (error) of which reads
\begin{equation}\label{eq:decomposition}
\err{A}{L \circ \Theta}{\rho}^{2}
	= \dst{A}{\Theta}{\rho}^{2} + \err{\pf{\Theta}{\rho}A}{L}{(\Theta\rho)}^{2}
\end{equation}
for any $A \in \lobspq{H}{\rho}$ and $\rho \in \stspq{H}$.  Given the fact \cite{Lee_2020_01, Lee_2020_01_Entropy} that every element of the space $\lobspq{H}{\rho}$ admits a (sequence of) measurement(s) that is capable of measuring it errorlessly (in the limit), the decomposition \eqref{eq:decomposition} allows for an operational characterization of the disturbance as the infimum
\begin{equation}\label{char:error}
\dst{A}{\Theta}{\rho}
	= \inf_{L} \err{A}{L \circ \Theta}{\rho}
\end{equation}
of the error of the composite measurement, along with the interpretation of the pushforward $\pf{\Theta}{\rho}A$ as the indicator of the locally optimal choice(s) of the (sequence of) secondary measurement(s) $L$ that attains the infimum (in the limit).  It goes without saying that, in general, for given $\stsp{S}_{1}$ and $\stsp{S}_{2}$, parallel characterization of the loss 
\begin{equation}
\loss{\ob{A}}{\process{T}_{1}}{s}
	= \inf_{\process{T}_{2}} \loss{\ob{A}}{\process{T}_{2} \circ \process{T}_{1}}{s}
\end{equation}
associated with a process $\process{T}_{1}: \stsp{S}_{1} \to \stsp{S}_{2}$ through a family of secondary processes $\{\process{T}_{2}: \stsp{S}_{2} \to \stsp{S}_{3}\}$ is available, granted the existence of a (sequence of) member(s) of the family that is capable of processing the local observable $\dpf{(\process{T}_{1})}{s} \ob{A} \in \lobsp{O}{s_{2}}$ losslessly over $s_{2}$ (in the limit).

The (de-\nobreak\hspace{0pt})\nobreak\hspace{0pt}composition law \eqref{eq:composition_law_process} reveals that our loss \eqref{def:loss_process} respects the (pre)order structure of `informativeness' of processes;  to be more explicit, given two processes $\process{T}_{1} : \stsp{S} \to \stsp{S}_{1}$ and $\process{T}_{2} : \stsp{S} \to \stsp{S}_{2}$, we may argue that the former is more informative than the latter given the existence of a process $\tilde{\process{T}}: \stsp{S}_{1} \to \stsp{S}_{2}$ with which the behavior of the latter can be fully described by the former as $\process{T}_{2} = \tilde{\process{T}} \circ \process{T}_{1}$, the situation of which may be written as $\process{T}_{1} \succeq \process{T}_{2}$.  Our definition of loss \eqref{def:loss_process} preserves this preorder as $\loss{\ob{A}}{\process{T}_{2}}{s} \geq \loss{\ob{A}}{\process{T}_{1}}{s}$, thereby revealing itself as an (reversed) order-homomorphism (order-preserving/isotone map).  Full discussion on this topic with its implications is beyond the scope of this paper, and shall thus be elaborated elsewhere.

\section{Uncertainty Relation for errors\label{sec:ur_joint}}

Prior to the introduction of our uncertainty relation for error and disturbance, we present a refined form of our previous relation \cite{Lee_2020_01,Lee_2020_01_Entropy} for quantum measurements involving errors, as it serves as a basis for the derivation of the former.  Let $M : \stspq{H} \to \stspc{\Omega_{1}}$ and $N : \stspq{H} \to \stspc{\Omega_{2}}$ be two quantum measurements chosen independently.   Suppose that the two measurements admit a joint description in the sense that there exists an affine map $J : \stspq{H} \to \stspc{\Omega_{1} \times \Omega_{2}}$ from which both the distributions $\Mrho$ and $\Nrho$ are retrieved as marginals from the distribution $\Jrho$.  More explicitly, this implies the classical processes $\pi_{1} : \stspc{\Omega_{1} \times \Omega_{2}} \to \stspc{\Omega_{1}}$ and $\pi_{2} : \stspc{\Omega_{1} \times \Omega_{2}} \to \stspc{\Omega_{2}}$ that project the joint probability distributions to their respective marginals
\begin{align}
(\pi_{1} p)(\omega_{1}) &\defeq \int_{\Omega_{2}} p(\omega_{1},\omega_{2})\, d\omega_{2} \label{def:projection_x}, \\
(\pi_{2} p)(\omega_{2}) &\defeq \int_{\Omega_{1}} p(\omega_{1},\omega_{2})\, d\omega_{1} \label{def:projection_y},
\end{align}
satisfying $M = \pi_{1} \circ J$ and $N = \pi_{2} \circ J$.  Given the fact that the adjoints of \eqref{def:projection_x} and \eqref{def:projection_y} respectively read $(\pi_{1}^{\prime}f)(x,y) = f(x)$ and $(\pi_{2}^{\prime}g)(x,y) = g(y)$, their pullbacks are isometries, \textit{i.e.}, $\lnorm{ f }{\pi_{1}p} = \lnorm{ \spb{\pi_{1}}f }{p}$ and $\lnorm{ g }{\pi_{2}p} = \lnorm{ \spb{\pi_{2}} g }{p}$, where we have adopted the abbreviations $\spf{\process{T}} = \pf{\process{T}}{s}$ and $\spb{\process{T}} = \pb{\process{T}}{s}$ in order to avoid clutter (in what follows, this simplified notation will be occasionally used, granted that there is no room for confusion).  In other words, this allows for the identification of the function spaces $\lobspc{\Omega_{1}}{\Mrho}$ and $\lobspc{\Omega_{2}}{\Nrho}$ regarding each of the measurements with their images under the pullbacks $\spb{\pi_{1}}$ and $\spb{\pi_{2}}$, which are in turn subspaces of the larger space $\lobspc{\Omega_{1}\times\Omega_{2}}{\Jrho}$ of the joint measurement.  

Now that we have equipped ourselves with the necessary concepts and facts, let us present our result.  Let $A, B \in \lobspq{H}{\rho}$ with $\rho \in \stspq{H}$.  Then, for any pair of quantum measurements $M$ and $N$ admitting a joint description, the inequality
\begin{equation}\label{ineq:uncert_proj}
\err{A}{M}{\rho}\, \err{B}{N}{\rho}
	\geq \sqrt{ \mathcal{R}^{2} + \mathcal{I}^{2} }
\end{equation}
holds, where
\begin{multline}\label{def:real}
\mathcal{R}
	\defeq \dacomm{A}{B}{\rho} - \linpr{ \pf{M}{\rho}A }{ \pf{M}{\rho}B }{\Mrho} \\
	- \linpr{ \pf{N}{\rho}A }{ \pf{N}{\rho}B }{\Nrho} + \linpr{ \pf{M}{\rho} A }{ \pf{N}{\rho} B }{\Jrho}
\end{multline}
and
\begin{align}\label{def:imaginary}
\mathcal{I}
	\defeq \dcomm{A}{B}{\rho} - \dcomm{ \pb{M}{\rho}\pf{M}{\rho}A }{B}{\rho} - \dcomm{ A }{ \pb{N}{\rho}\pf{N}{\rho}B }{\rho}
\end{align}
with the commutator $[A,B] \defeq AB - BA$.  Here, we have also introduced the abbreviated notation
\begin{equation}\label{def:product_joint}
\linpr{ \pf{M}{\rho} A }{ \pf{N}{\rho} B }{\Jrho}
	\defeq \linpr{ \spb{\pi_{1}} \pf{M}{\rho} A }{ \spb{\pi_{2}} \pf{N}{\rho} B }{\Jrho}
\end{equation}
under the identifications $f \simeq \spb{\pi_{1}}f$ and $g \simeq \spb{\pi_{2}}g$ mentioned above.  

\begin{figure}
\includegraphics[hiresbb,clip,width=0.9\linewidth,keepaspectratio,pagebox=artbox]{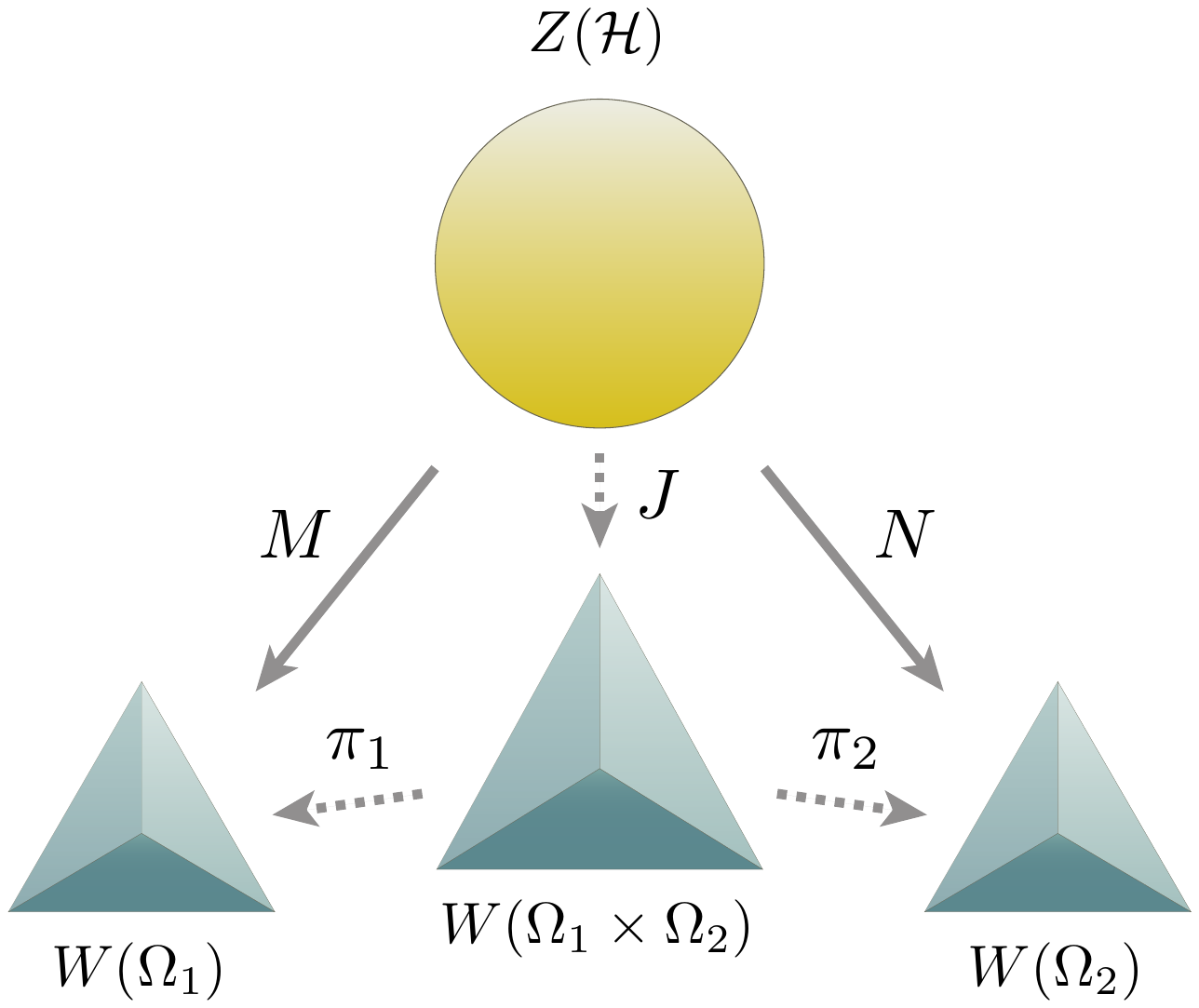}%
\caption{The basic structure of measurements leading to the uncertainty relation for errors associated with two quantum measurements $M$ and $N$.  The cause of the uncertainty lies in the presumption that there exists a quantum measurement $J$ that jointly describes the given two measurements $M$ and $N$.}
\label{figure3}
\end{figure}

The proof of the inequality \eqref{ineq:uncert_proj} is actually quite simple: it is just a direct corollary to the Cauchy--Schwarz inequality.  A quick way to see this is to first introduce the semi-inner product
\begin{equation}
\inpr{ (X, f) }{ (Y, g) }
	\defeq \lev{ X^{\dagger}Y }{\rho} + \lev{ f^{\dagger} g }{\Jrho} 
	- \lev{ J^{\prime}f^{\dagger} J^{\prime}g }{\rho}
\end{equation}
defined on the product space of Hilbert space operators and complex functions, as well as the seminorm $p(X,f) \defeq \sqrt{\inpr{ (X, f) }{ (X, f) }}$ that it induces.  Noticing the equalities
$\err{A}{M}{\rho} = p(X_{A}, f_{A})$ and $\err{B}{N}{\rho} = p(Y_{B}, g_{B})$,
where
$X_{A} \defeq A - \pb{M}{\rho}\pf{M}{\rho} A$,
$f_{A} \defeq \spb{\pi_{1}} \pf{M}{\rho} A$,
$Y_{B} \defeq B - \pb{N}{\rho}\pf{N}{\rho} B$,
and $g_{B} \defeq \spb{\pi_{2}} \pf{N}{\rho} B$,
we find that the Cauchy--Schwarz inequality for the product of $p(X_{A}, f_{A})$ and $p(Y_{B}, g_{B})$ becomes
\begin{equation}\label{ineq:cs}
\err{A}{M}{\rho}\, \err{B}{N}{\rho}
	\geq \abs{ \inpr{ (X_{A}, f_{A}) }{ (Y_{B}, g_{B}) } }.
\end{equation}
The semi-inner product appearing in the right-hand side of \eqref{ineq:cs} is a complex number $\inpr{ (X_{A}, f_{A}) }{ (Y_{B}, g_{B}) } = \mathcal{R} + i\,\mathcal{I}$ whose real part $\mathcal{R}$ and the imaginary part $\mathcal{I}$ are respectively given by \eqref{def:real} and \eqref{def:imaginary}.  This completes our proof of the inequality \eqref{ineq:uncert_proj}.

The terms $\mathcal{R}$ and $\mathcal{I}$ bear their own meanings; the former is a bound that is shared in common with the classical case (\lq semiclassical bound\rq), whereas the latter provides an additional contribution to the former preexisting bound, which is characteristic of quantum measurements (\lq quantum bound\rq).  This is supported by the observation that, if the lower bound is considered for the product of the loss \eqref{def:loss_process} of a pair of classical processes $K_{i} : \stspc{\Omega} \to \stspc{\Omega_{i}}$, $i =1,2$, a parallel inequality to \eqref{ineq:uncert_proj} can be derived through analogous arguments (needless to say, such an inequality can be derived for general processes in the same way), for which $\mathcal{I}$ vanishes and only the real part $\mathcal{R}$ contributes to the bound.

Before proceeding further, we stress that behind the appearance of the lower bound lies the existence of a joint distribution map $J$ for the two given measurements $M$ and $N$, which is not at all taken for granted in general.   An elementary situation where this is the case is when $M$ and $N$ are projection measurements \eqref{def:projection_measurement} respectively associated with observables $\hat{M}$ and $\hat{N}$ that are commutative $[\hat{M},\hat{N}] = 0$ as operators.
In such a case, the quantum measurement $J : \rho \mapsto (\Jrho)(m_{i},n_{j}) = \tr{\vert m_{i}\rangle\langle m_{i} \vert \vert n_{j}\rangle\langle n_{j} \vert \rho}$ induced by combining the spectral decompositions $\hat{M} = \sum_{i} m_{i}\, \vert m_{i}\rangle\langle m_{i} \vert$ and $\hat{N} = \sum_{j} n_{i}\, \vert n_{j}\rangle\langle n_{j} \vert$ of the two operators provides a simple example of the joint distribution map.

Although the current form of presentation suffices for our main purpose, we note that the condition of joint describability of two quantum measurements $M$ and $N$ admits obvious generalization; the essence is the existence of a quantum measurement $J: \stspq{H} \to \stspc{\Omega}$ mediating the behaviors of the two measurements $M$ and $N$ with classical processes $\pi_{i}:\stspc{\Omega} \to \stspc{\Omega_{i}}$, $i = 1,2$, \textit{i.e.}, $M = \pi_{1} \circ J$ and $N = \pi_{2} \circ J$.  If the pullbacks of both classical processes $\pi_{i}$ happen to be linear isometries over a certain state, which is always the case globally when both $\pi_{i}$ are projections to the marginals as exemplified above, the same line of arguments leads to the same uncertainty relation as \eqref{ineq:uncert_proj} valid over the said state.  One of the elementary situations in which this is the case is when one measures two observables with a single measurement $M$; this amounts to the special case $M = N$ which admits the trivial joint description by $J = M$ and trivial classical processes $\pi_{i} = \Id$, which subsequently yields the uncertainty relation for errors given in \cite{Lee_2020_01,Lee_2020_01_Entropy}.  For general cases (in which the pullbacks of $\pi_{i}$ are not necessarily isometries), the uncertainty relation still holds in the form of \eqref{ineq:uncert_proj} with due modification to the term $\mathcal{R}$.

A proper and comprehensive treatment of such general cases necessitates more elaborate mathematical tools, which is beyond the scope of this paper, and will thus be given elsewhere; there the potential non-uniqueness of the classical bound $\mathcal{R}$, which is dependent on the generally non-unique choice of the map $J$, will also be investigated.  In view of this, we may obviously take the supremum of the term $\abs{ \mathcal{R} }$ for all such possible $J$ of our inequality \eqref{ineq:uncert_proj}, or just consider the simplified version
\begin{equation}\label{ineq:uncert_proj_simple}
\err{A}{M}{\rho}\, \err{B}{N}{\rho}
	\geq \abs{ \mathcal{I} },
\end{equation}
which is free from the choice of $J$ and contains only the quantum bound;  needless to say, the same structures and remarks also apply to our uncertainty relation \eqref{ineq:uncert_error_disturbance} for error and disturbance as well.

\section{Uncertainty Relation for Error and Disturbance\label{sec:ur}}

We are now ready to introduce our uncertainty relation for error and disturbance.  Let $M : \stspq{H} \to \stspc{\Omega_{1}}$ be a quantum measurement and let $\Theta_{M} : \stspq{H} \to \stspq{K}$ denote its observation effect, \textit{i.e.}, the quantum process the measurement $M$ inevitably causes on the system $\hsp{H}$ to be measured.  Here, we stress again that our process $\Theta_{M}$ need not be confined to completely positive maps that end in the same quantum system $\hsp{K} = \hsp{H}$ as the initial system, the assumption of which many formulations including Ozawa's \cite{Ozawa_2003} necessitate;  this allows for the description of the measurement process as well as the quantitative evaluation of the resultant disturbance that were previously not quite viable, \textit{e.g.}, measurement through high energy collision involving particle decays, thereby entertaining deeper implications regarding the conception of the observation effect than are commonly conceived.  

\begin{figure}
\includegraphics[hiresbb,clip,width=0.9\linewidth,keepaspectratio,pagebox=artbox]{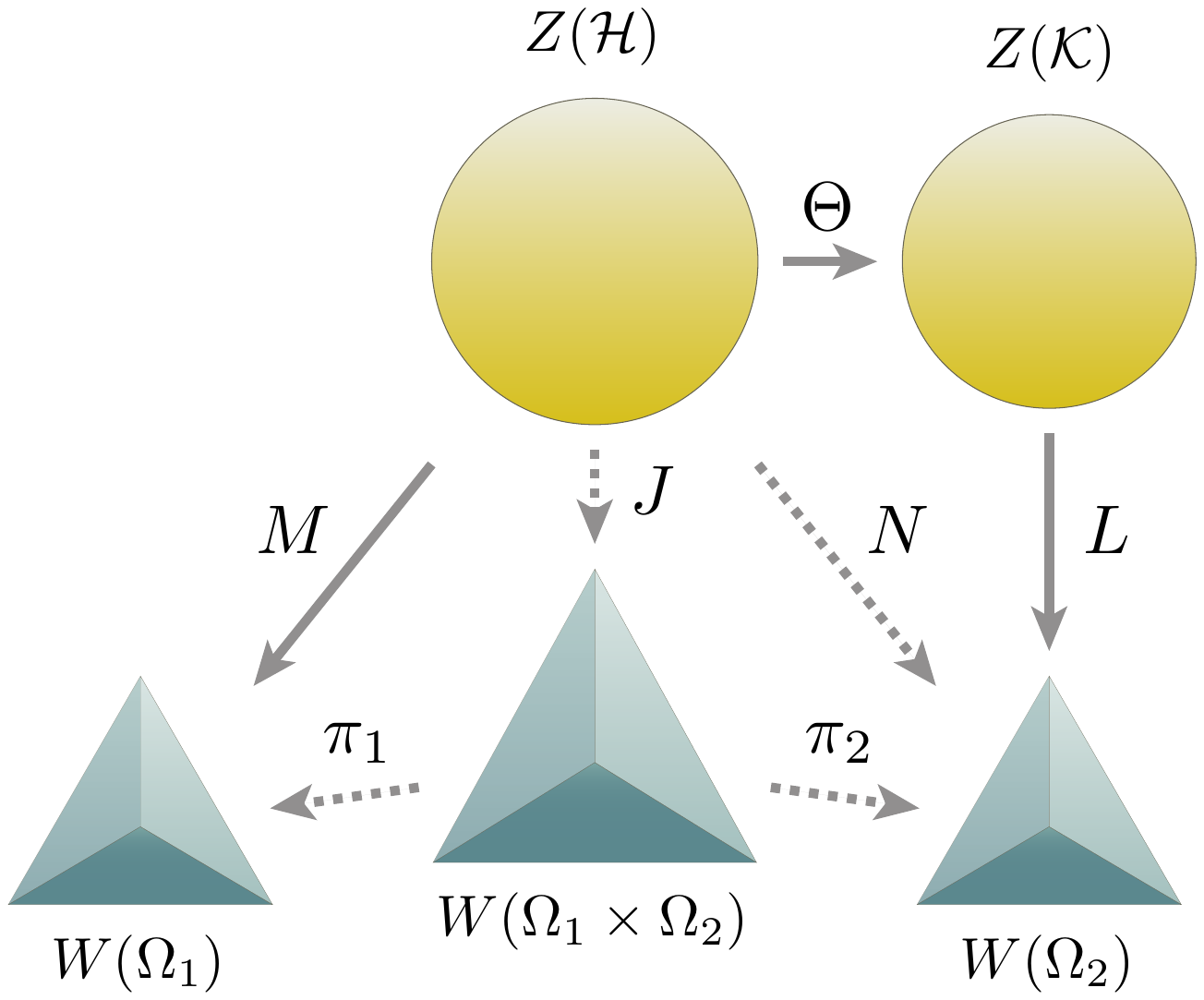}%
\caption{The structure of generating error and disturbance associated with a measurement.  The primary measurement, which is meant to measure the observable $A$ in the system $\stspq{H}$, induces a map $M$ to the space of probability distributions $\stspc{\Omega_{1}}$ describing its measurement outcomes.  
At the same time, the measurement induces a map $\Theta$ to the space of quantum states $\stspq{K}$ describing the resultant 
states caused by the effect of the measurement.   The secondly measurement, which is meant to measure the observable $B$ in the system $\stspq{K}$,  induces a map $L$ to the space of probability distributions $\stspc{\Omega_{2}}$ describing its measurement outcomes. The map $M$ and the composite map $L \circ \Theta$ provide the pair of maps $M$ and $N$ and their joint distribution 
$\stspc{\Omega_{1} \times \Omega_{2}}$ discussed before (see FIG. \ref{figure3}). }
\label{figure4}
\end{figure}

Let us now consider a sequential measurement in which the primary measurement $M$ is followed by a secondary measurement $L$;  the secondary measurement is performed after the system underwent the process $\Theta_{M}$ induced by $M$, the corresponding map of which is given by $L : \stspq{K} \to \stspc{\Omega_{2}}$.    Viewed from the original state space $\stspq{H}$, this sequential measurement by $M$ and $L$ naturally defines a joint measurement $J : \stspq{H} \to \stspc{\Omega_{1} \times \Omega_{2}}$ describing both the primary $M = \pi_{1} \circ J$ and the secondary $L \circ \Theta_{M} = \pi_{2} \circ J$ measurements with the projections \eqref{def:projection_x} and \eqref{def:projection_y} introduced earlier.  It is important to note that the quantum process $\Theta_{M}$ must be constrained strictly by the measurement map $M$ in order for the composite process $L \circ \Theta_{M}$ to possess a joint measurement $J$ along with the primary measurement $M$ for arbitrary $L$, which is by no means always the case for general $\Theta$. 

In general, let $M$ and $\Theta$ be any pair of a quantum measurement and process on $\hsp{H}$ for which the above property holds \textit{i.e.}, there always exists a joint measurement $J$ of $M$ and $N = L \circ \Theta$ for any quantum measurement $L$ on $\hsp{K}$.  At this point, we may choose $L$ in such a way that the disturbance of the observable $B$ on $\hsp{H}$ caused by the process $\Theta$ coincides
\begin{equation}\label{ineqed:cs}
\dst{B}{\Theta}{\rho}
	= \err{B}{L \circ \Theta}{\rho}
\end{equation}
with the error of the composite measurement (in the sense of limit, if necessary).
Indeed, we have seen previously in \eqref{char:error} that this choice is not only possible, but also supported from the ground that it is the infimum of the error of the composite measurement $L \circ \Theta$ for all possible $L$.  Then, by substituting $N = L \circ \Theta$ in \eqref{ineq:uncert_proj}, one immediately obtains the inequality
\begin{equation}\label{ineq:uncert_error_disturbance}
\err{A}{M}{\rho}\, \dst{B}{\Theta}{\rho}
	\geq \sqrt{ \mathcal{R}^{2} + \mathcal{I}^{2} }
\end{equation}
for error and disturbance, with $\mathcal{R}$ and $\mathcal{I}$ being respectively given by \eqref{def:real} and \eqref{def:imaginary} with due substitutions for $N$.  In particular, for the case where $\Theta = \Theta_{M}$ is the quantum process representing the observation effect of $M$, the relation \eqref{ineq:uncert_error_disturbance} admits an interpretation as the uncertainty relation of the error of a measurement and the disturbance of its observation effect.

To be more explicit, the semi-classical and quantum contributions to the lower bound respectively reads
\begin{multline}\label{def:real_error_disturbance}
\mathcal{R} 
	= \dacomm{A}{B}{\rho} - \linpr{ \pf{M}{\rho}A }{ \pf{M}{\rho}B }{\Mrho} \\
	- \linpr{ \pf{\Theta}{\rho}A }{ \pf{\Theta}{\rho}B }{\,\Theta\rho} + \linpr{ \pf{M}{\rho} A }{ \pf{\Theta}{\rho} B }{\Jrho}
\end{multline}
and
\begin{align}\label{def:imaginary_error_disturbance}
\mathcal{I}
	= \dcomm{A}{B}{\rho} - \dcomm{\pb{M}{\rho}\pf{M}{\rho}A}{B}{\rho} - \dcomm{A}{\pb{\Theta}{\rho}\pf{\Theta}{\rho}B}{\rho},
\end{align}
where we have introduced the shorthand
$\linpr{ \pf{M}{\rho} A }{ \pf{\Theta}{\rho} B }{\Jrho}
	\defeq \linpr{ \spb{\pi_{1}} \pf{M}{\rho} A }{ \spb{\pi_{2}} \pf{N}{\rho} B }{\Jrho}$
under the identifications in the similar vein as \eqref{def:product_joint}.  Here, note the equivalence (in the sense of limit, if need be) of the condition \eqref{ineqed:cs} to $\err{\pf{\Theta}{\rho}B}{L}{(\Theta\rho)} = 0$ by \eqref{eq:decomposition}, which is in turn equivalent to $\spf{\Theta}B = \spb{L}\spf{L} (\spf{\Theta}B)$ as shown in \cite{Lee_2020_01,Lee_2020_01_Entropy}, whereby $\spb{N}\spf{N} B = \spb{(L \circ \Theta)} \spf{(L \circ \Theta)} B = \spb{\Theta} \spb{L}\spf{L}\spf{\Theta} B = \spb{\Theta}\spf{\Theta}B$ follows from the composition laws of the pullback \eqref{eq:composition_pullback} and the pushforward \eqref{eq:composition_pushforward}.  This allows us to rewrite the quantum bound \eqref{def:imaginary} into \eqref{def:imaginary_error_disturbance}, which is found to be independent of the choice of $L$ as expected.

\section{Affinity with Heisenberg's Original Idea\label{sec:affinity}}

Just as with our uncertainty relation for errors \cite{Lee_2020_01,Lee_2020_01_Entropy} and its refinement \eqref{ineq:uncert_proj}, our relation for error and disturbance \eqref{ineq:uncert_error_disturbance} also implies a potential violation of the traditional bound $\abs{ \lev{ [A, B] }{\rho} / 2i }$ for certain choices of quantum measurements and their observation effects.  It is to be emphasized, however, that even though the product of the error and the disturbance may violate the traditional bound quantitatively, Heisenberg's original idea of the uncertainty principle remains valid; errorless measurement of an observable $A$ is impossible without disturbing another observable $B$ whenever $\lev{ [A, B] }{\rho} \neq 0$.  We shall now argue why this is the case.

In general, the situation in which the loss \eqref{def:loss_process} of a process vanishes $\loss{\ob{A}}{\process{T}}{s} = 0$ shall be referred to as being \emph{lossless};  as for the two specific cases, lossless quantum measurements (Q-C processes) and lossless quantum processes (Q-Q processes) shall also respectively be referred to as \emph{errorless} \cite{Lee_2020_01,Lee_2020_01_Entropy} and \emph{non-disturbing}.  Some of the characterizations of the lossless process are:
\begin{enumerate}[label=\rm{(\alph*)}]
\item $\loss{\ob{A}}{\process{T}}{s} = 0$,\label{fact:lossless-process_1}
\item $\ob{A} = \pb{\process{T}}{s}\dpf{\process{T}}{s}\, \ob{A}$,\label{fact:lossless-process_2}
\item $\lnorm{ \ob{A} }{s} = \lnorm{ \pb{\process{T}}{s}\dpf{\process{T}}{s} \mathcal{A} }{s}$.\label{fact:lossless-process_3}
\end{enumerate}
Indeed, \ref{fact:lossless-process_2} $\implies$ \ref{fact:lossless-process_3} is trivial, \ref{fact:lossless-process_3} $\implies$ \ref{fact:lossless-process_1} is a direct consequence of the non-expansiveness $\lnorm{ \dpf{\process{T}}{s}\, \mathcal{A} }{\process{T}(s)} \geq \lnorm{ \pb{\process{T}}{s}\dpf{\process{T}}{s} \mathcal{A} }{s}$ of the pullback, and \ref{fact:lossless-process_1} $\implies$ \ref{fact:lossless-process_2} follows from the evaluation $\loss{\ob{A}}{\process{T}}{s} \geq \lnorm{ \mathcal{A} - \pb{\process{T}}{s}\dpf{\process{T}}{s} \mathcal{A} }{s}$.  For quantum measurements, the conditions have been spelled out in Refs.~\cite{Lee_2020_01,Lee_2020_01_Entropy}, whereas for quantum processes, they explicitly read: 
\begin{enumerate*}[label=\rm{(\alph*)}]
\item $\dst{A}{\Theta}{\rho}= 0$,\label{fact:non-disturbing-process_1}
\item $A = \pb{\Theta}{\rho}\pf{\Theta}{\rho} A$\label{fact:non-disturbing-process_2}, and
\item $\lnorm{ A }{\rho} = \lnorm{ \pb{\Theta}{\rho}\pf{\Theta}{\rho} A }{\rho}$.\label{fact:non-disturbing-process_3}
\end{enumerate*}

A direct corollary to our uncertainty relation \eqref{ineq:uncert_error_disturbance} is: given $\lev{ [A, B] }{\rho} \neq 0$ for a pair $A, B \in \lobspq{H}{\rho}$, there is no quantum measurement $M$ over $\rho$ that is capable of measuring $A$ or $B$ errorlessly without causing disturbance to the other.  Indeed, if there were such a measurement, the relation combined with the equivalence $\ref{fact:non-disturbing-process_1} \iff \ref{fact:non-disturbing-process_2}$ above would immediately lead us to a contradiction $0 \geq  \sqrt{\abs{ 0 }^{2} + \abs{ \lev{ [A, B] }{\rho}/ 2 }^{2} } > 0$, which may be readily confirmed by plugging $\pb{M}{\rho}\pf{M}{\rho}A = A$ and $\pb{\Theta}{\rho}\pf{\Theta}{\rho}B = B$ (or $\pb{M}{\rho}\pf{M}{\rho}B = B$ and $\pb{\Theta}{\rho}\pf{\Theta}{\rho}A = A$, depending on the choice of the observables concerned) into the lower bound to find $\mathcal{R} = 0$ while $\mathcal{I} = - \lev{ [A, B] }{\rho}/(2i)$ in \eqref{ineq:uncert_error_disturbance}. One thus finds that for non-trivial (\textit{i.e.}, $\mathrm{dim}(\hsp{H}) \geq 2$) quantum systems, there exists no quantum measurement without observation effect.  

Note, however, that our formulation does not necessarily prohibit either of the error or disturbance from vanishing.  For instance, in view of the fact that $\err{A}{M}{\rho} = 0$ implies $\mathcal{R} = 0$ and $\mathcal{I} = - \lev{ [A,\pb{\Theta}{\rho}\pf{\Theta}{\rho}B] }{\rho} / (2i) = 0$, it is thus obvious that an errorless measurement imposes a heavy constraint on $\Theta$;  still, as long as the latter constraint is satisfied, the relation \eqref{ineq:uncert_error_disturbance} yields no contradiction.

One may convince oneself with a simple example that this may indeed be the case:  consider the projection measurement \eqref{def:projection_measurement} associated with a non-degenerate observable $A$, for which we have the spectral decomposition $\hat{M} = A = \sum_{i=1}^{N} a_{i}\, |a_{i}\rangle\langle a_{i}|$.  One may straightforwardly confirm that this measurement furnishes an errorless measurement $\err{A}{M}{\rho} = 0$ of $A$, whereas, regarding its observation effect $\Theta = \Theta_{M}$, the traditional projection postulate \eqref{def:collapse} entails $\lev{ [A, \pb{\Theta}{\rho} C] }{\rho} = 0$ for all $C \in \lobspq{H}{\rho}$ through a straightforward computation, thereby (by putting $C = \pf{\Theta}{\rho}B$) satisfying the constraint.

Analogously, the uncertainty relation \eqref{ineq:uncert_proj} for joint measurements forbids errorless measurement of a pair $A, B \in \lobspq{H}{\rho}$ with $\lev{ [A, B] }{\rho} \neq 0$, even if the measurements concerned are (as long as they admit a joint description) not necessarily identical $M \neq N$ (\textit{cf}. Ref.~\cite{Lee_2020_01, Lee_2020_01_Entropy}).  This points to the fact that, for non-trivial quantum systems, there exists no (pair of) quantum measurement(s) (admitting a joint description) that is capable of measuring every observable errorlessly over every state.

The relation \eqref{ineq:uncert_proj} for joint measurements does not prohibit either of the errors from vanishing as well.  For instance, observe that $\err{A}{M}{\rho} = 0$, which is realised by the projection measurement $M$ associated with $\hat{M} = A$, necessitates the constraint $\lev{ [A,\pb{N}{\rho}\pf{N}{\rho}B] }{\rho} = 0$.  Meanwhile, for any projection measurement $N$ associated with $\hat{N}$ satisfying $[\hat{M},\hat{N}] = 0$ as operators, which guarantees the joint measurability of $M$ and $N$, the constraint is readily confirmed to be satisfied.

\section{Conclusion and Discussions\label{sec:discussion}}

In this paper, we have presented a general uncertainty relation for error and disturbance based on our universal formulation, which is experimentally verifiable without any conditions for operational implementability, while maintaining Heisenberg's original idea of the uncertainty principle, albeit in a more involved manner than is usually recognized.  Concerning this, one of the most notable benefits of adopting our universal formulation is that it reveals the cause of quantum uncertainty more clearly: it is the existence of the mediating (joint) distribution that constrains the possible form of state transformations induced by the measurement.

Given the universality and generality of our framework, it should be also worthwhile to consider whether our results can shed some light on other notable uncertainty relations mentioned in the Introduction.  In view of this, our relation for errors embraces the standard Kennard--Robertson (Schr{\"o}dinger) relation regarding quantum indeterminacy expressed by standard deviations as its special case when the measurement is non-informative \cite{Lee_2020_01,Lee_2020_01_Entropy}.  As for the relation for errors, Ozawa's relation \cite{Ozawa_2004_01} as well as those of Arthurs, Kelly, and Goodman \cite{Arthurs_1965, Arthurs_1988}, are also understood to be corollaries to the special cases of ours (see Refs.~\cite{Lee_2020_01,Lee_2020_01_Entropy}).

Regarding the relation for the observation effect, we first note that our framework naturally encompasses the indirect measurement scheme adopted by several alternative formulations including Ozawa's, for every quantum measurement employing detector (meter) systems also preserves the structure of probabilistic mixture.  We find it assuring in this respect that our relation reduces Ozawa's relation \cite{Ozawa_2003}, which especially assumes quantum measurements $M: \stspq{H} \to \stspc{\mathbb{R}}$ with real $\Omega = \mathbb{R}$ outcomes, and that their observation effects $\Theta_{M} : \stspq{H} \to \stspq{H}$ are completely positive maps between identical $\hsp{H} = \hsp{K}$ quantum systems, to one of the corollaries to its special cases;  under such assumptions, our inequality is tighter than his.  Indeed, whenever Ozawa's definitions \cite{Ozawa_2003} of error (noise, in his terminology) $\varepsilon$ and disturbance $\eta$ are well-defined, so are ours $\varepsilon_{\rho}$ and $\eta_{\rho}$, and the former are respectively never less than the latter;  this may simply be observed from the fact that they admit expressions as special cases of our $\mathcal{B}$-loss \eqref{def:loss_gauge}, namely that of a Q-C process $M$ with $\mathcal{B}$ being the identity map on the real line, and that of a Q-Q process $\Theta_{M}$ with $\mathcal{B} = \mathcal{A}$, respectively.
One then further reveals $\varepsilon(A) \eta(B) \geq \varepsilon_{\rho}(A) \eta_{\rho}(B) \geq \sqrt{\mathcal{R}^2 + \mathcal{I}^2} \geq \abs{ \mathcal{I} } \geq \abs{ \lev{ [A,B] }{\rho} } /2 - \varepsilon(A)\sigma(B) - \sigma(A)\eta(B)$, the left- and right-most hand sides of which is equivalent to Ozawa's inequality.  More details on these topics, including the mathematical subtleties behind the formulations, shall be reported in our subsequent papers.

Our universal formulation enables us to see that the three orthodox relations regarding quantum indeterminacy, measurement error, and observation effect, each of which governs the seemingly distinct realms in which the uncertainty principle manifests itself, are in fact parallel manifestations of a single principle, sharing common conceptual and mathematical structures.

\begin{acknowledgments}
The authors thank Prof.~Naomichi Hatano for fruitful discussions and insightful comments.  This work was supported by JSPS Grant-in-Aid for Scientific Research (KAKENHI), Grant Numbers JP18K13468 and JP18H03466.
\end{acknowledgments}

\bibliography{}

\end{document}